\documentclass[aip,apl,reprint,superscriptaddress,showpacs]{revtex4-1}

\usepackage{graphicx}
\usepackage{color}
\definecolor{darkgreen}{RGB}{34, 139, 34}
\usepackage{dcolumn}
\usepackage{bm}
\usepackage{amsmath,amssymb}

\renewcommand{\d}{{\rm d}}
\newcommand{\e}{{\rm e}}
\newcommand{\imai}{{\rm i}}

\begin{document}

\title{Nonlinear response of quantum cascade structures}

\author{David O. Winge, Martin Lindskog, and Andreas Wacker}
\email[]{Electronic mail: Andreas.Wacker@fysik.lu.se}
\affiliation{Mathematical Physics, Lund University, Box 118, 22100 Lund, Sweden}

\date{31. October 2012, accepted by Applied Physics Letters}

\begin{abstract} 
The gain spectrum of a terahertz quantum cascade laser is analyzed by a non equilibrium Green's functions approach.
Higher harmonics of the response function were retrievable, providing a way to approach nonlinear phenomena in quantum cascade lasers theoretically. 
Gain is simulated under operation conditions and results are presented both for linear response and strong laser fields. 
An iterative way of reconstructing the field strength inside the laser cavity at lasing conditions is described using a
measured value of the level of the losses of the studied system. 
Comparison with recent experimental data from time-domain-spectroscopy indicates that the experimental
situation is beyond linear response.
\end{abstract}

\maketitle

Possible coherent radiation in the terahertz range  has been a very
strong motivation for research in the field of Terahertz Quantum
Cascade Lasers\cite{KohlerNature2002,WilliamsNatPh2007} (THz-QCLs),
which would enable a wide range of applications  such as
imaging\cite{DarmoOE2004} and spectroscopy\cite{HubersAPL2006}.
However compact devices operating over cryogenic temperatures are a
practical requirement for applications and currently the most
promising designs are based on resonant phonon extraction
\cite{WilliamsAPL2003},  achieving operating temperatures up to 
$\sim$~200 K.\cite{FathololoumiOE2012}

The key physical quantity in any QCL is the gain which describes the
amplification of the optical field in the heterostructure material. In
recent years this quantity has been measured in detail in
Time-Domain-Spectroscopy (TDS)
experiments\cite{KrollNature2007,BurghoffAPL2011} where THz-QCLs are
probed by ultra short pulses providing information on both phase and
amplitude of the transmitted pulse, whereafter the gain spectrum is
reconstructed by a Fourier transform. The pulse is made as strong as
possible in order to get a good signal to noise ratio but it is not
known how the system dynamics are affected by such a measurement. The
simulation of THz QCLs relies on a consistent treatment of tunneling
and scattering, either by hybrid density matrix/rate equation schemes
\cite{CallebautJAP2005,KumarPRB2009,TerazziNJP2010,DupontPRB2010,BhattacharyaAPL2012}
or more evolved Non Equilibrium Green's Function (NEGF) theory
\cite{LeePRB2006,SchmielauAPL2009,KubisPRB2009,HaldasIEEE2011}. Here
we present an extension of our NEGF scheme\cite{WackerSPIE2009}
towards the treatment of high intensities inside the QCL, going beyond
linear response to an external electromagnetic field.

In this article we consider a time-dependent electric field
$F(t)=F_{\rm dc}+F_{\rm ac}\cos(\Omega t)$ in the cavity, which reflects both
the applied bias ($F_{\rm dc}$) and the electric component ($F_{\rm ac}$) of a
monochromatic field of angular frequency $\Omega$ in the cavity.
Going beyond linear response, this requires the solution of the time-dependent
Kadanoff-Baym equation for the lesser and
retarded Green's functions, 
$G^<_{\alpha\beta}({\bf k},t_1,t_2)$ and
$G^{\rm ret}_{\alpha\beta}({\bf k},t_1,t_2)$, respectively
\cite{HaugJauhoBook1996}. Here $\alpha,\beta$ denote the states in growth
direction and ${\bf k}$ the in-plane momentum. 
The periodicity in time allows for a Fourier 
decomposition of the Green's functions
\begin{equation}
G({\bf k};t_1,t_2)=\frac{1}{2\pi}\int \d E\sum_h \,\e^{-\imai
  E(t_1-t_2)/\hbar} G_h({\bf k},E) \e^{-\imai h \Omega t_1}
\label{EqGreens}
\end{equation}
and similarly for the self-energies. This provides a set of equations for the
Green's functions for given self-energies $\Sigma_{\alpha\beta, h}({\bf
    k},E)$, which are defined analogously. This procedure follows essentially
the concepts outlined in Ref.~\onlinecite{Brandes_PRB1997} and details will be
given elsewhere.
Here the terms with $h=0$ correspond to the stationary transport considered
before \cite{LeePRB2006}, while the higher order terms take 
into account the ac field.
For the fields considered in this manuscript we used $h=-2,-1,0,1,2$, while
checking that increasing $|h|$ did not change the results (generally higher values of $|h|$ are required with increasing ratios 
$\e F_{\rm  ac}d/\hbar \Omega$, where $d$ is the period of the QCL structure).
Relations to observables are made through the $h=0$ and $h=1$ components although the higher orders
effect the lower ones implicitly. The Green's functions allow for a determination of the current, where
$G^<_{\alpha\beta\, h}({\bf k},E)$ provides the dc current for $h=0$ and the 
ac current with frequency $\Omega$ for $h=1$. Dividing the latter by $F_{\rm
  ac}$ provides the conductivity, directly related to the gain coefficient.

\begin{figure}
\includegraphics[width=0.9\columnwidth]{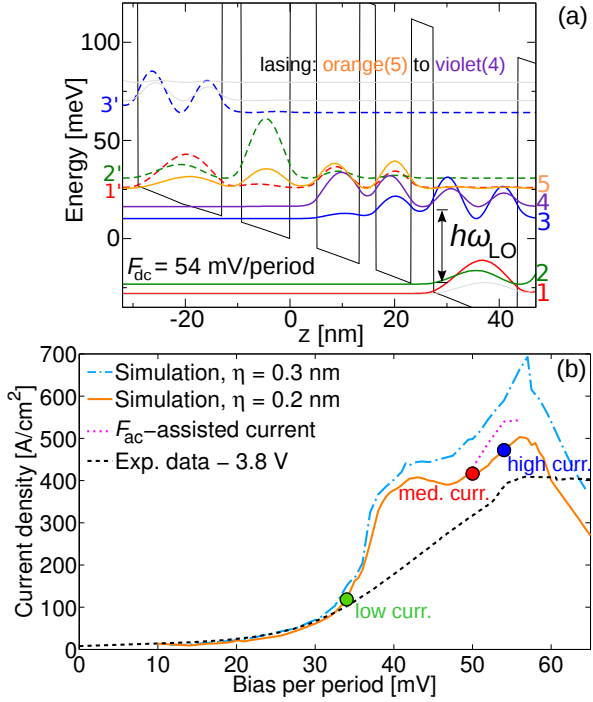}
\caption{(Color online) (a) QCL structure considered with the main states contributing to its operation and (b) 
calculated current-bias relation for $\eta = 0.2$ nm (solid) and $\eta = 0.3$ nm (dotted-dashed) together with experimental 
data of Ref.~\onlinecite{BurghoffAPL2011} (dashed). The shift in current at simulated operating conditions (dotted) is discussed later. 
The marked points at different current densities are analyzed in Fig.~\ref{FigGain}.}
\label{FigSample}
\end{figure}

\begin{figure}
\includegraphics[width=\columnwidth]{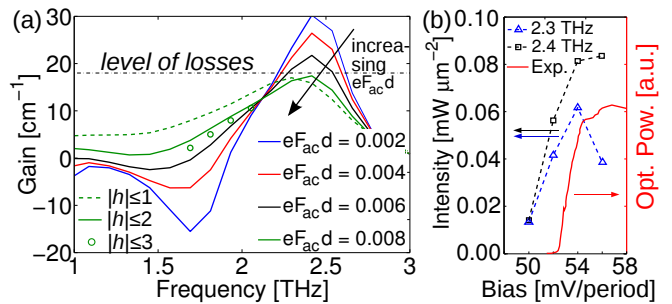}
\caption{(Color online) (a) Gain spectra at a bias of 54 mV/period at different ac field strengths. (b) Calculated intensities in the waveguide (dashed) and corresponding experimental data (full line).}
\label{FigClamping}
\end{figure}

Here we consider the sample studied in Ref.~\onlinecite{BurghoffAPL2011} as
shown in Fig.~\ref{FigSample}(a). In Fig.~\ref{FigSample}(b) we show the
current-bias relation, calculated by our model, together with experimental
data. The original experimental data refer to the bias $U$ along the entire
heterostructure, containing 175 periods of length $d$ as well as contact
regions, where additional bias drops. In a similar
experiment\cite{BurghoffAPL2012} a voltage drop of around 3.0 V as well as a
5$\Omega$ series resistance were assumed for the data analysis. Here, we use a
voltage drop of 3.8 V in the contacts for converting the experimental bias to
$eF_{\rm dc}d$ as displayed in Fig.~\ref{FigSample}(b) and
Fig.~\ref{FigClamping}(b).  

Fig.~\ref{FigSample}(b) displays simulations with different interface roughness scattering parameters in order to calibrate one of the 
parameters used, namely the average (RMS) of the roughness height $\eta$. This enters in the matrix element
for the interface roughness scattering self energy\cite{LeePRB2002} together with the typical size $\lambda=10$ nm of the roughness layers. 
According to Fig.~\ref{FigSample}(b) it is clear that a lower value of $\eta$ suppresses the current flow, and regarded as a fit 
parameter $\eta = 0.20$ nm is a better value.\footnote{A higher values of
 $\eta$ also reduces  gain below  18 cm$^{-1}$, which would  prevent from
  lasing operation, not shown} All subsequent simulations where carried out
with $\eta = 0.20$ nm and a lattice temperature of 77~K entering the
occupation of the phonon modes. The experimental
heat-sink  temperature was $\approx 33$ K \cite{BurghoffAPL2011}, but  the 
lattice temperature for resonant phonon extraction THz-QCLs is typically
higher\cite{VitielloAPL2005} due to heating effects.

We note that for both simulations the main peak as well as
the low-bias behavior are in good agreement with the experimental data.
In contrast, for bias drops per period of about 40-45 mV, we observe
a spurious extra peak due to tunneling between state 1 and state 3 of the next neighboring period.
Such extra peaks for long-range tunneling have been observed in our 
model for other THz-structures as well, and we currently attribute them to
the lack of electron-electron scattering processes\footnote{This possibility 
had been pointed out to us by H. Callebaut and Q. Hu}.

\begin{figure}
\includegraphics[width=\columnwidth]{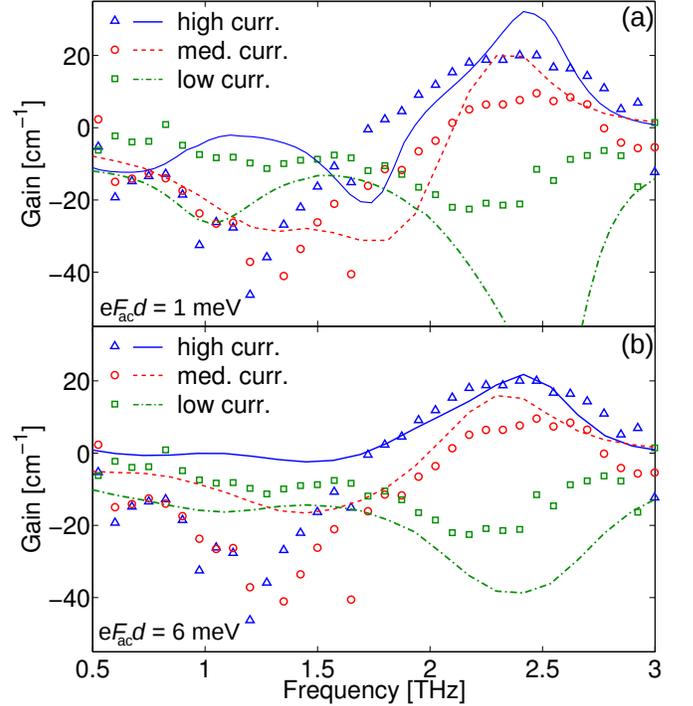}
\caption{(Color online) Gain spectrum at different operation conditions for 
small $\e F_{\rm ac}d=1$~meV (a) and larger $\e F_{\rm ac}d=6$~meV (b) ac field strength. 
Open symbols are experimental data, and lines are simulated data. 
Green (dotted-dashed) corresponds to simulations at 34 mV/period, red (dashed)
to 50 mV/period and blue (solid) to 54 mV/period. The experimental currents were 63, 319 and 403 A/cm$^2$ for 
the low, medium and high current respectively.}
\label{FigGain}
\end{figure}

Fig.~\ref{FigGain} shows the calculated gain spectra (lines) for three
different dc biases, as depicted in Fig.~\ref{FigSample}(b), both for low (a) and high (b) ac field strength. 
For comparison, we display the experimental data from Ref.~\onlinecite{BurghoffAPL2011}
in both parts.
 The low current point ({\color{darkgreen} {\footnotesize $\square$}/-$\cdot$-}) is considered to be at the bias where state 1' and 4 become
aligned and electrons start to tunnel through the system as those levels are
at resonance.  
Here we are far below threshold and mainly absorption is seen
at the laser frequency of 2.2~THz, as state 4 -- the lower laser state -- is
populated, while 5 is still mostly empty. This is changed as current is
increased and we approach the medium current point ({\color{red}
  $\circ$/-\,-}). 
This is taken where the system is almost at, but still below the threshold
current. Here  the states 1' and 5 start to  align, creating population inversion
at the laser frequency. At operating conditions, where the third and last point ({\color{blue} {\footnotesize$\triangle$}/-}) at high current is taken, 
gain is above the level of the losses\cite{BurghoffAPL2011} 
of 18~cm$^{-1}$ and can now sustain lasing as the population inversion is at
its maximum.  

The simulations at low and high ac field strength shown in Fig.~\ref{FigGain}
differ drastically, but the general picture is that around the laser
frequency, large losses at low current develop into gain as bias is raised, as
observed by Refs.~\onlinecite{MartlOE2011,BurghoffAPL2011}. At a more detailed
level, the low ac field strength simulations exhibit stronger features which
are not reflected in the experimental data. At higher ac field strength
however, the overall agreement becomes better, mainly due to a redistribution
of carriers (bleaching).  However, the strong absorption  feature around 1.2
THz for medium and high current density in the experimental data does not
appear in the simulations, for which we do not have any explanation.  The
better agreement of experimental data with high ac field, indicates  that the
experimental conditions are beyond linear response.

Here it is important to address the fundamental differences between experiment
and simulation. In the simulations, a monochromatic ac field is applied and
the gain spectrum is constructed frequency by frequency. In the experimental
case, the situation is quite different. A pulse,  containing all frequencies
within the bandwidth (typically 3~THz\cite{KrollNature2007}) is sent into the
sample, and the way this pulse has changed by passing through the structure
determines the gain spectrum. Compared to the simulations, where we only
measure at the frequency where we excite, this is the opposite, as all
frequencies are  subject to excitations and all frequencies are also
measured. Therefore, our modeling can only be seen as approximate.  In
addition, the experimental data is the difference of measurements between the
unbiased QCL and the QCL at the chosen measuring bias. This way the background
is effectively subtracted. If the structure exhibits  less losses at some
point than it does at zero bias, this is measured as gain. In the simulations
however, we only extract the gain from the conductivity  extracted from the
$h=1$ component of the Green's functions, as we do not have to take losses
into account and thus only look at the \textit{intrinsic} gain spectra.
Simulations at very low bias, i.e. the off-state, show absorption peaks at 
0.9~THz and 2.7~THz, which could explain corresponding features in the
experimental gain spectra.

We have demonstrated, that the response varies significantly with the ac field
strength. Thus  it is important to question whether the ac field strengths
used in the simulations are comparable to their experimental counterpart. In
order for the effects of high ac field strengths shown in
Fig.~\ref{FigGain}(b) to be of any relevance, the power coupled into the QCL
structure during the experimental measurements must be sufficient. Addressing
this question, consider the experimental situation governing the in-coupling
of light\cite{BurghoffAPL2011}: A pumping femtosecond pulse of 125 mW  hits
the emitter section of the same QCL as described in Fig.~\ref{FigSample}. The
pulse generates a photocurrent giving an electric field  transient that is
coupled across an air distance of 4 $\mu$m into the QCL section of interest.
Our value $\e F_{\rm ac} d= 6$ meV corresponds to a power of 40 mW in the
cavity, which requires an extremely efficient conversion in the emitter and
good coupling  between the structures. It is far from clear, whether the
probing field can reach these intensities. Strong ac fields at lasing
conditions would be capable of generating these effects, but this would then
only contribute above threshold current.  

   In a working laser, the gain will clamp at the level of the
losses, as the population inversion will stabilize around the configuration
where the inversion lost to the  optical field will be balanced by the
injector efficiency. This happens at different intensities for different bias
points, and using this fact the intensity at  gain clamping and thus the power
in the QCL can be reconstructed theoretically.  One can use the level of the
losses measured in Ref.~\onlinecite{BurghoffAPL2011} in order to iteratively
calculate the  intensity at operating conditions. Fig.~\ref{FigClamping}(a)
shows gain spectra at a bias of 54 mV/period that have been simulated at
various  ac-field strengths. It is clear from Fig.~\ref{FigClamping} that
higher ac field strengths effectively lower both gain and absorption, and
give rise to gain bleaching.  For all bias points above threshold simulations
were carried out at different ac field strengths in order to see which ones
gave gain matching the level of the losses at 18 cm$^{-1}$. The corresponding
intensities were then extracted and are shown in
Fig.~\ref{FigClamping}(b). For the maximum intensity of 
0.08~mW/$\mu$m$^{2}$ the wave guide area of 800~$\mu$m$^2$ provides 
a power of 64~mW inside the QCL. This seems reasonable
taking into account that only a part is coupled out
through the mirror. 

To show the importance of including higher orders of the Fourier decomposed
Green's function in Eq.~(\ref{EqGreens}), simulations with $|h| \leq 1$ only and
also $|h|\leq3$ is shown in Fig.~\ref{FigClamping}(a) for $\e F_{\rm ac} d= 8$
meV which is the highest ac-field used.  $|h| \leq 1$ (dashed) shows a clear
deviation from the $|h| \leq 2$ case (full line) while the simulations with
$|h| \leq 3$ confirm the quality of our $|h| \leq 2$ calculations for 
$\e F_{\rm  ac}d \lesssim /\hbar \Omega$.

The $F_{\rm ac}$-assisted current at the intensities shown in
Fig.~\ref{FigClamping}(b) is displayed as a dotted line in
Fig.~\ref{FigSample}(b). It increases proportional to the intensity compared
to the non-lasing current. Thus when lasing sets in, we see a kink in the
current, just as in the experimental data of Fig.~\ref{FigSample}(b) at 53~mV
per period. As the calculated kink appears somewhat stronger, our calculated
lasing intensities could be a little bit too high. This may be related to the
fact, that the experimental lasing frequency of 2.2~THz is not precisely at
the peak of the gain spectrum.

In conclusion, we have simulated gain under operation by including higher
orders of the Fourier decomposed Green's function in order to include
nonlinear effects.  We have found a way to calculate the power of the laser
using an experimental value of the level of the losses and by iteratively
matching the gain to that level and then extract the intensity of such a
configuration. It has also been shown that gain bleaches under high intensity
conditions and that  this might be a non negligible effect in THz-TDS
measurements.  

\begin{acknowledgments}
We thank Dayan Ban for helpful discussions and providing the experimental
data of Ref.~\onlinecite{BurghoffAPL2011}.
Financial support from the Swedish Research Council (VR) is gratefully acknowledged.
\end{acknowledgments}

\end{document}